\documentstyle[aps,prl,epsf]{revtex}
\begin{document}
\draft

\wideabs{
\title{SiC(0001): a surface Mott-Hubbard insulator}
\author{V.I. Anisimov$^1$, A.E. Bedin$^1$, M.A. Korotin$^1$,
G. Santoro$^{2,3}$, S. Scandolo$^{2,3}$, E. Tosatti$^{2,3,4}$}
\address{$^1$Institute of Metal Physics, Ekaterinburg, GSP-170, Russia\\
$^2$International School for Advanced Studies (SISSA), Via Beirut 2,
Trieste, Italy\\
$^3$Istituto Nazionale per la Fisica della Materia (INFM), Via Beirut 2, 
Trieste, Italy\\
$^4$International Centre for Theoretical Physics (ICTP), Trieste, Italy}
\date{\today}
\maketitle

\begin{abstract}
We present ab-initio electronic structure calculations for
the Si-terminated SiC(0001)$\sqrt{3}\times\sqrt{3}$ surface. 
While local density approximation (LDA) calculations
predict a metallic ground state with a half-filled narrow band, Coulomb
effects, included by the spin-polarized LDA+U method, result in a 
magnetic (Mott-Hubbard) insulator with a gap of $1.5$ eV, 
comparable with the experimental value of $2.0$ eV.
The calculated value of the inter-site exchange parameter, $J=30$K, leads to
the prediction of a paramagnetic Mott state, except at
very low temperatures. The observed Si 2p surface core level
doublet can naturally be explained as an on-site exchange splitting.
\end{abstract}
}                

\noindent
The fractional adlayer structures on semiconductors show rich 
phase diagrams with potential instabilities of charge density 
wave (CDW) and spin density wave (SDW) type 
at low temperatures, as well as Mott insulating phases \cite{santoro}.
The (0001)$\sqrt{3}\times \sqrt{3}$ surface of hexagonal SiC,
(as well as the closely analogous (111)$\sqrt{3}\times \sqrt{3}$ of 
cubic SiC) expected from standard LDA calculations to be a 2D metal with 
a half-filled narrow band of surface states in the bulk energy gap
\cite{SiC-struc,pollmann} --
is instead experimentally proven to be an insulator with a rather large 
($2.0$ eV) band gap \cite{johansson,IPES}.
It has been suggested that this system is a Mott 
insulator\cite{neugebauer} due to the large value of ratio $U/W$, 
where $U$ is the Coulomb interaction parameter, of the order of several eV, 
and $W$ is the surface bandwidth, calculated to be about $0.35$ eV
\cite{SiC-struc,pollmann}. 
Very recent STM data further confirm this picture \cite{rama}. 
However, a detailed electronic description of the ensuing state 
and of its magnetic implications is 
not yet available.

Here we present a novel electronic structure calculation for this 
system, based on the LSDA+U method \cite{ldau1}, which is able to 
take into account Coulomb interactions between localized electrons. 
The result -- an insulating surface, with a large intra-adatom exchange
splitting, but an exceedingly weak antiferromagnetic inter-adatom 
exchange coupling -- 
is now in quantitative agreement with available photoemission data. 

The generalized LSDA+U method\cite{lichtan} is devised to partially 
cure the weakness of the local spin density approximation 
(LSDA) functional in dealing with interacting electrons in strongly 
localized orbitals, by supplementing the standard LSDA functional, 
$E^{\rm LSDA}[\rho^\sigma ({\bf r)}]$ where $\sigma=\uparrow,\downarrow$, 
with a mean-field Hartree-Fock factorization, $E^U$, 
of the screened Coulomb interaction $V_{\rm ee}$ among electrons
in localized atomic orbitals $i,m$. 
(Here $i$ collectively labels the quantum number $n$, the angular momentum $l$, 
and the atomic site, while $m$ labels the projection of the angular momentum.)
The generalized LSDA+U functional is then written as:
\begin{equation}  \label{E_LSDAU:eqn}
E^{\rm LSDA+U}=E^{\rm LSDA}[\rho^\sigma ({\bf r)}] 
+ \sum_{i} E^U[\{n^{i,\sigma}\}] - E_{\rm dc} \;,
\end{equation}
where $n^{i,\sigma}=n^{i,\sigma}_{mm'}$ is the localized orbital occupation 
density matrix. 
Double counting of the localized orbital contribution to the energy, 
already included by LSDA in an average way, is taken into account, 
in Eq.\ (\ref{E_LSDAU:eqn}), by subtracting a compensating term
\begin{equation}  \label{dc:eqn}
E_{\rm dc}=\frac 12 \sum_i
[ UN_i(N_i-1)-J\sum_{\sigma}N_i^{\sigma}(N_i^{\sigma}-1) ],
\end{equation}
where $U$ and $J$ are the screened Coulomb and exchange parameters  
\cite{superlsda,anigun},
$N_i^\sigma=\sum_m n_{mm}^{i,\sigma}$, 
and $N_i=N_i^{\uparrow}+N_i^{\downarrow}$. 

Minimization of this modified functional with respect to the charge density
$\rho^{\sigma}({\bf r})$ and the orbital occupation $n^{i,\sigma}_{mm'}$
leads to an effective single particle Hamiltonian 
\begin{equation}  \label{hamilt}
\widehat{H}=\widehat{H}_{\rm LSDA}+
\sum_{i,\sigma}\sum_{mm^{\prime}} \mid im\sigma \rangle
V_{mm^{\prime}}^{i,\sigma} \langle im^{\prime}\sigma \mid
\end{equation}
which corrects the usual LSDA potential through a localized 
orbital contribution of the form:
\begin{eqnarray}  \label{Pot}
V_{mm^{\prime}}^{i,\sigma} &=& \sum_{\{m\}}
\{ \langle m,m^{\prime\prime}
\mid V_{\rm ee} \mid m^{\prime},m^{\prime\prime\prime}\rangle 
( n_{m^{\prime\prime}m^{\prime\prime\prime}}^{i,\sigma}
+ n_{m^{\prime\prime}m^{\prime\prime\prime}}^{i,-\sigma} )
\nonumber \\
&& -\langle m,m^{\prime\prime} \mid V_{\rm ee} \mid 
m^{\prime\prime\prime},m^{\prime} \rangle 
n_{m^{\prime\prime}m^{\prime\prime\prime}}^{i,\sigma} \} 
\nonumber \\
&&-U(N_i-\frac 12)+J(N_i^{\sigma}-\frac 12) \;.
\end{eqnarray}
Important ingredients in the calculation are the matrix elements of 
screened Coulomb interaction $V_{\rm ee}$ which we parameterize,
in analogy with the atomic case, in terms of effective Slater integrals 
$F^k$ \cite{JUDD} which, in turn, can be linked to the 
Coulomb and Stoner parameters $U$ and $J$, as obtained from 
LSDA-supercell constrained calculations \cite{anigun,ANISOL}.
For Si, we apply the above corrections to $3p$ states only, since $3s$ states 
are fully occupied. The calculated Coulomb and Stoner 
parameters for Si $3p$ states are $U=8.6$ eV and $J=1$ eV, respectively.
The electronic structure calculations we have performed rely on 
the Linear Muffin-Tin Orbitals (LMTO) method \cite{lmto}, and use 
the Stuttgart TBLMTO-47 code. An important point is that the 
calculation allows us, through the Green's
function method, to evaluate the effective inter-site exchange interaction 
parameters $J_{ij}$, as second derivatives of the ground state energy 
with respect to the magnetic moment rotation 
angle \cite{lichtexchange,lichtan}.

\begin{figure}
\centerline{\epsfxsize=3.0in\epsfysize=3.0in %
\epsfbox{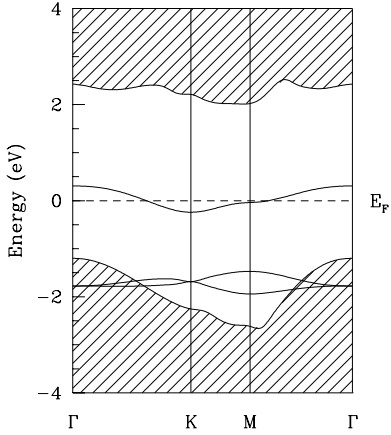}} 
\caption{
Energy bands obtained from a standard LDA calculation for the
SiC(0001)$\protect\sqrt{3}\times\protect\sqrt{3}$ surface. 
Brillouin zone notations correspond to a
unit cell with one Si adatom. Energy is measured from the Fermi level.
Note the narrow half-filled surface band.
}
\label{lda:fig}
\end{figure}

The SiC(0001)$\sqrt{3}\times\sqrt{3}$ surface was simulated by a slab
containing two SiC bilayers plus a Si adatom layer. 
The Si adatoms were placed in the $T_4$ positions of the upper (Si) surface. 
All atomic positions were fixed to the values calculated by Ref.\ 
\cite{SiC-struc}. 
The bottom surface (C atoms of fourth atomic layer)
was saturated with hydrogen atoms. A standard LDA calculation 
for the $\sqrt{3}\times \sqrt{3}$  unit cell containing one Si adatom gave, 
as expected, 
a metallic ground state with a half-filled narrow band situated within the 
energy gap of bulk SiC (Fig.\ \ref{lda:fig}). 
This surface band originates from the dangling bonds of the Si adatoms 
($p_z$-orbitals). The corresponding Wannier function, however, contains 
less than $50\%$ of the adatom $p_z$ orbitals, for, 
as is well-known \cite{SiC-struc}, it extends itself down into the first 
SiC bilayer.

The LSDA+U method, relying on a Hartree-Fock-like mean-field approximation, 
can only deal with statically long-range ordered ground states. 
It is generally believed\cite{noncol} that a triangular lattice 
of electrons at commensurate filling, with nearest neighbor hopping, and a 
strong Hubbard repulsion -- i.e., in the Heisenberg limit --
possesses, in spite of strong quantum fluctuations, a three-sublattice 
120$^{\circ}$-N\'eel long-range order 
-- a commensurate spiral spin density wave with 120$^{\circ}$ spins 
lying on a plane.
This kind of non-collinear magnetic order can in principle be handled
by LSDA+U, through a straightforward extension of the functional. 
However, that extension makes the scheme not only computationally 
heavier, but of intrinsically worse quality than the corresponding 
collinear calculation, as recently found in another case \cite{gebauer}.
We therefore decided to carry out our calculation in the correct 
three-sublattice supercell (with three Si adatoms) but to restrict 
to collinear magnetic moments. 
Non-collinearity is then taken into account within a much
less computationally heavy Hubbard model, the resulting small value
of antiferromagnetic coupling further supporting the superior
accuracy of this approach.

\begin{figure}
\centerline{\epsfxsize=3.5in\epsfysize=3.0in %
\epsfbox{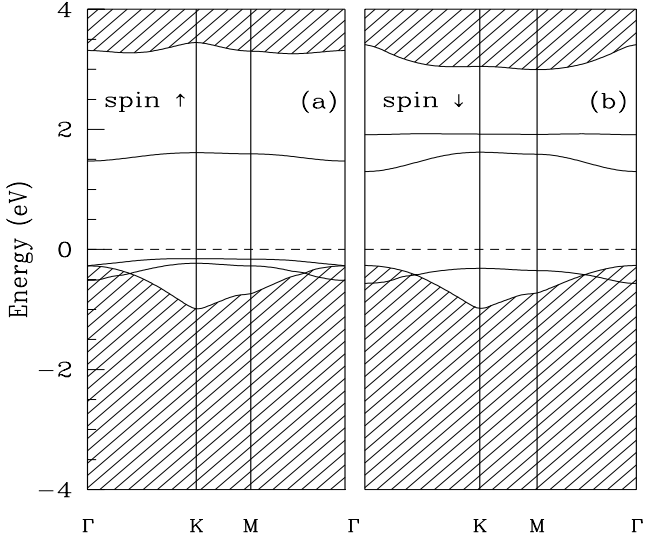}} 
\caption{
LSDA+U energy bands for the 
spin-polarized collinear state of SiC(0001)$3\times 3$ surface. Brillouin zone
notations correspond to the unit cell with three Si adatoms. (a): spin-up
electrons, (b): spin-down electrons. Energy is measured from the Fermi level.
Note the insulating state, with a gap of about 1.5 eV. 
}
\label{bands:fig}
\end{figure}

We started our LSDA+U calculation assuming two adatoms with a finite
spin-up projection of the magnetic moment, and the third with spin-down.
The result converges to a stable magnetic 
insulating ground state with an energy gap of $1.5$ eV, which compares rather
well with the experimental value of $2.0$ eV \cite{IPES}. 
The corresponding energy bands are shown in Fig.\ \ref{bands:fig}.
Of the three spin-up bands, two are occupied and one is empty,
(Fig.\ \ref{bands:fig}(a)), 
whereas for spin-down electrons (Fig.\ \ref{bands:fig}(b)) one band is 
occupied and two are empty. Consequently, there is a net magnetization
corresponding to one spin for three adatoms. Of course, this
net magnetic moment will disappear in the true noncollinear
antiferromagnetic state,
(except perhaps for a small fraction which might be stabilized by surface
spin-orbit coupling).

\begin{figure}
\centerline{\epsfxsize=3.0in\epsfysize=3.0in %
\epsfbox{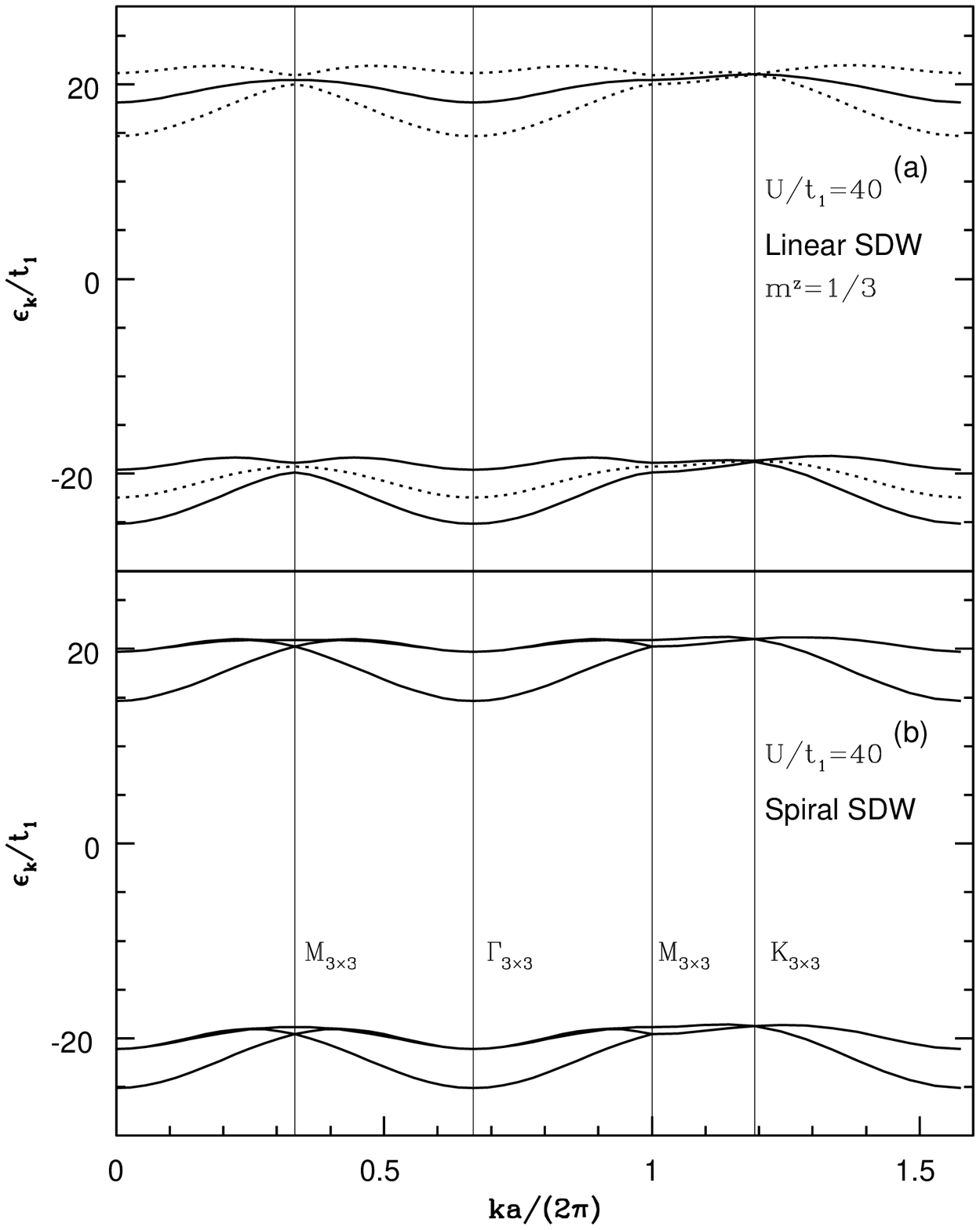}} 
\caption{
Tight-binding HF electronic bands along high symmetry directions of the
Brillouin Zone for the one-band Hubbard model at 
$U_{\rm eff}/t_1=40$ for the two magnetic solutions: 
(a) linear SDW with a net uniform magnetization $m^z=1/3$;
(b) spiral SDW, with finite on-site magnetization, and  $m^z=0$. 
Solid and dashed lines denote up and down bands, respectively.
}
\label{ek_u40:fig}
\end{figure}

As a further check that the restriction to collinear spins does not
really alter the main features of the bands in a substantial way, we
performed a Hubbard model calculation for the possible
$3\times 3$ magnetic phases, within a Hartree-Fock (HF) 
treatment \cite{santoro}. 
Restricting ourselves to a one-band model with a nearest-neighbor
hopping $t\approx 0.04$ eV (as extracted from the LDA surface bandwidth), 
we use an effective Hubbard $U$ roughly equal to the ab-initio energy 
gap of $\approx 1.5$ eV. 
The results for the two magnetic solutions found are shown in Fig.\
\ref{ek_u40:fig}(a,b). 
Fig.\ \ref{ek_u40:fig}(a) shows the linear SDW solution with a uniform
magnetization $m^z=1/3$, which compares rather well with the ab-initio
surface bands. 
Allowing for non-collinear spins, we then obtain the spiral SDW solution
whose bands are shown in Fig.\ \ref{ek_u40:fig}(b). 
We stress the fact that the spiral SDW is the actual HF ground state, 
while the linear SDW is only a metastable state. We note, however,
that the ground state energies of the two solutions differ by less 
than $0.5$ meV/adatom, so that unrealistically precise ab-initio calculations
would be necessary to decide which of the two is the actual ground state. 
Moreover, the HF bands for the two solutions are rather similar,
apart from some extra splittings introduced by the collinear magnetic solution.
It is not clear whether the present resolution of photoemission data
\cite{johansson,IPES} would allow to distinguish between these two 
slightly different sets of bands.

Coming back to the ab-initio study, we have calculated a magnetic moment 
for the Si adatom of approximately $0.6\mu_B$. This reduction is of
course not due to quantum fluctuations, absent in the LSDA+U, but to
the subsurface delocalization of the Wannier function,
whose weight is only about half on the Si adatom.
For the same reason, the energy gap value of $1.5$ eV is 
roughly $4-5$ times smaller than the bare $U$ value.
In fact, the LSDA+U correction to the potential of a particular orbital 
$m\sigma$ is roughly proportional to $U(1/2-n_{m\sigma})$, and the
splitting of the potentials for spin-up and spin-down $p_z$-orbitals of Si
adatoms is proportional to the magnetic moment on the adatom. 
This means that this potential splitting can be estimated to be roughly $U/2$. 
If one takes into account that the potential correction is applied to 
Si$3p$-orbital, which is 
about one half of the total Wannier function corresponding to 
the half-filled band, its effect on the energy splitting $\Delta E$ will 
be further reduced by a factor two, yielding roughly $\Delta E\approx U/4$, 
close to the value obtained in our actual LSDA+U calculation.

We have also calculated
the value of the inter-site exchange 
parameter $J_{ij}$ corresponding to the interaction between the spins of the
electrons localized on neighboring Si adatoms, and obtained a value 
of $J_{ij}=30$K. The corresponding one-band Hubbard model 
estimate is $J_{ij}=4t^2/U_{\rm eff}$, where $t$ is the 
nearest-neighbor hopping parameter and $U_{\rm eff}$ is the effective Coulomb 
Hubbard $U$. Taking $U_{\rm eff}$ of the order of the energy gap 
($\approx 2$ eV) and $t=0.04$ eV, we obtain $37$K, which agrees 
quite well with the LSDA+U result. 

The existence of finite magnetic moments on the Si adatoms should be 
experimentally detectable. Methods which require magnetic long-range
order do not seem viable, since the only circumstance
where a finite-temperature order parameter could survive
in this frustrated 2D magnetic system, would be in presence of a 
large spin-orbit induced magnetic anisotropy favoring out-of-surface ordering, 
which we do not have reason to expect. 
Therefore, without ruling out the possibility of low-temperature ordering, we 
believe that the SiC(0001)$\sqrt{3}\times \sqrt{3}$ 
surface will be in an overall paramagnetic Mott insulating state,
in spite of the existence of an on-site moment, at least down to liquid 
nitrogen temperature.

\begin{figure}
\centerline{\epsfxsize=3.0in\epsfysize=3.5in %
\epsfbox{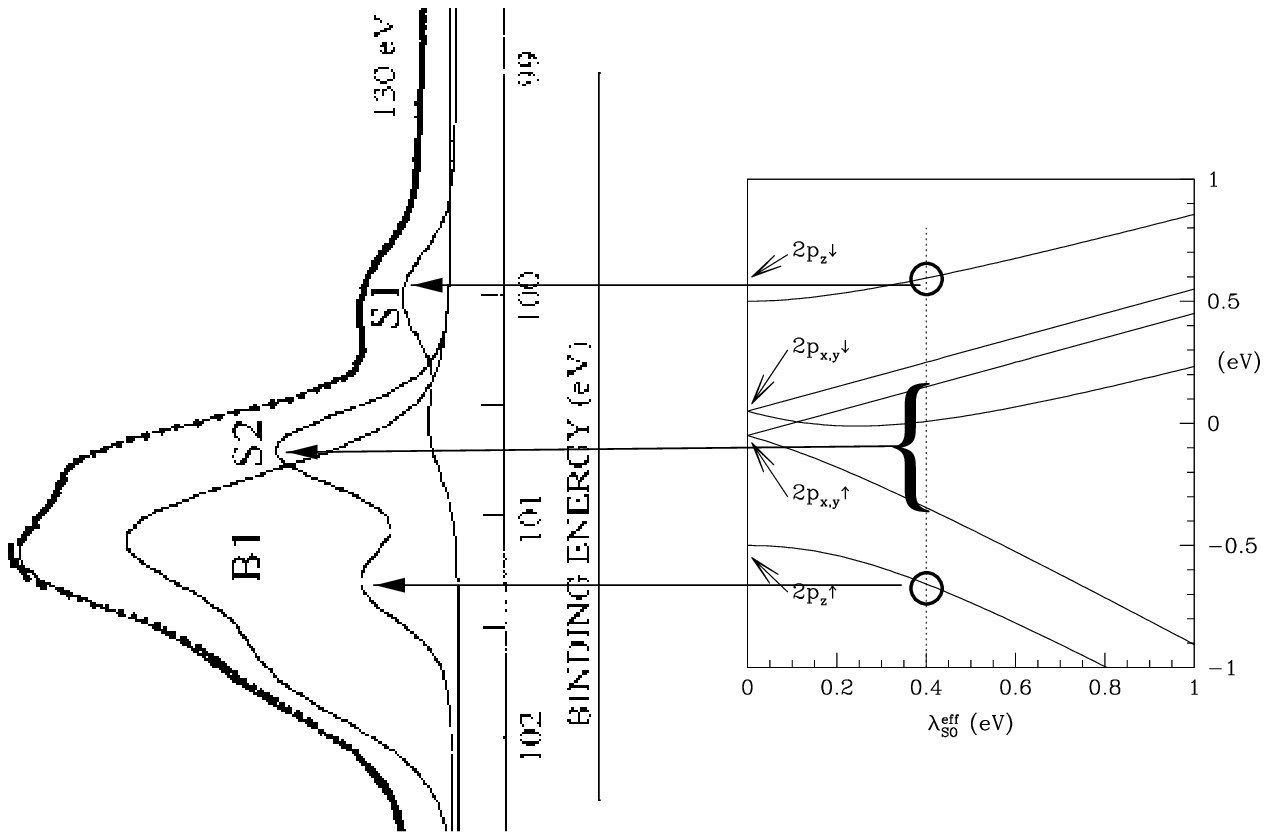}} 
\caption{
Si 2p core levels calculated as a function of spin-orbit coupling,
in comparison with photoemission data and original fitting in terms
of bulk (B1) and two surface sites S1 and S2 by Johansson 
\protect{\it et al.\/}\protect\cite{core}. 
Upon inclusion of intra-atomic exchange splitting the 
whole surface contribution S1+S2 can be explained as due to a single site. 
}
\label{levels:fig}
\end{figure}

The situation is much more promising at the intra-atomic level. 
Here, exchange splittings in the Si adatom 2p core levels should be large 
and detectable. One complication is that the adatom dangling 
bond has a predominant $3p_z$ character, which
breaks the core level symmetry between $2p_z$ and $2p_{x,y}$ by an amount
not negligible in comparison to spin-orbit coupling and exchange. 
The $2p$ core hole will exhibit a $6$-fold multiplet resulting from
the joint effect of intra-atomic exchange, asymmetry, and spin-orbit. 
An estimate for the multiplet splitting is obtained by assuming the 
valence electrons to be distributed with weights $\alpha$ in 
$3p_z\uparrow$, and $\beta/4$ in each $3p_{x,y}\uparrow\downarrow$. 
For $\lambda_{\rm SO}=0$, we calculate a bare asymmetry splitting
$\Delta=\epsilon_{2p_x\uparrow}-\epsilon_{2p_z\uparrow}= 
\alpha [(J_{zz}-J_{zx})-(K_{zz}-K_{zx})] 
+ (\beta/4) [2(K_{zz}-K_{zx})-(J_{zz}-J_{zx})]$, where
$K_{zz(x)} = 
\left( 2p_{z(x)}, 3p_z | e^2/|{\bf r}_1-{\bf r}_2| | 2p_{z(x)}, 3p_z \right)$, 
and 
$J_{zz(x)} = 
\left( 2p_{z(x)}, 3p_z | e^2/|{\bf r}_1-{\bf r}_2| | 3p_z, 2p_{z(x)} \right)$, 
are Coulomb and exchange integrals. 
The first square bracket dominates, and is estimated by a $Si^+$ Hartree-Fock
calculation to be about $1$ eV, so that $\Delta \approx \alpha$ (eV).
Since only about $50\%$ of the dangling bond orbital is $3p_z$, we obtain
the desired crude estimate $\Delta \approx 0.5$ eV.
Exchange splittings are large for the $2p_z$ case 
and small for the $2p_{x,y}$, 
reflecting the large differences in the corresponding
exchange integrals $J_{zz}$ and $J_{zx}$. 
When we include the spin-orbit interaction 
$\lambda_{\rm SO} {\bf L} \cdot {\bf S}$, the final core hole levels
are obtained as in Fig.\ \ref{levels:fig}. 
For $\lambda_{\rm SO}=0.4$ eV we find a roughly three-peaked structure,
the broad central peak four times as strong as each side peak.
This offers an alternative explanation of the experimental lineshape\cite{core}
(see Fig.\ \ref{levels:fig}) in terms of a single exchange-split multiplet,
rather than two chemically inequivalent sites $S_1$ and $S_2$ \cite{core} 
whose existence is otherwise not supported. 

Additional direct experimental evidence for the Mott-Hubbard state
could be obtained by a careful study of adatom ($3p_z$, $3p_z$)
Auger spectral intensities, which should be easily singled out owing
to the large gaps.
The probability of double occupancy of the adatom dangling bond orbital 
should be almost completely suppressed, dropping from the band value 
of $1/4$, to a value 
of order $t/U$ which is one order of magnitude smaller. 
Even considering that only half of the orbital
is adatom $3p_z$, this surface should show 
a ($3p_z$, $3p_z$) Auger intensity which, by comparison
with the remaining (3p,3p) values, is anomalously small, 
as a proof of its Mott-Hubbard state.

More experimental and theoretical effort is
clearly called for to check these strong correlations and 
related magnetic effects, 
possibly the first of this magnitude to be suggested 
for an sp bonded, valence semiconductor surface. 

We thank S. Modesti for useful discussions.
Work at SISSA was partially supported by INFM through PRA LOTUS and HTSC,
by MURST through COFIN97, 
and by the EU, through contracts ERBCHRXCT940438 and FULPROP ERBFMRXCT970155.
Work at IMP was supported by the Russian Foundation for Basic
Research grant RFFI-98-02-17275.


\begin{center} {\bf Figures} \end{center}


\begin{references}

\bibitem{santoro} 
G. Santoro {\it et al.\/}, Surf. Sci. {\bf 402-404}, 802 (1998); 
G. Santoro, S. Scandolo, E. Tosatti, Phys.\ Rev.\ B {\bf 59}, 1891 (1999). 

\bibitem{SiC-struc}  
J.E.Northrup and J.Neugebauer, Phys.\ Rev.\ B {\bf 52}, R17001 (1995)

\bibitem{pollmann} 
M. Sabisch, P. Kruger, and J. Pollmann, Phys.\ Rev.\ B {\bf 55}, 10561 (1997).

\bibitem{johansson}
L. I. Johansson {\em et al.\/}, Surf.\ Sci.\ {\bf 360}, L478 (1996);

\bibitem{IPES}  
J.-M. Themlin {\it et al.\/}, Europhys.\ Lett.\ {\bf 39}, 61 (1997). 

\bibitem{neugebauer} 
J.E. Northrup and J. Neugebauer, Phys.\ Rev.\ B {\bf 57}, R4230 (1998).

\bibitem{rama} 
V. Ramachandran and R.M. Feenstra, Phys.\ Rev.\ Lett.\ {\bf 82}, 1000
(1999).

\bibitem{ldau1}  
V.I. Anisimov, J. Zaanen and O.K. Andersen, 
Phys.\ Rev.\ B {\bf 44}, 943 (1991); 
V.I. Anisimov, F. Aryasetiawan, A.I. Lichtenstein,
J. Phys.: Condens.\ Matter {\bf 9}, 767 (1997).

\bibitem{lichtan}  
A.I. Lichtenstein, J. Zaanen, V.I. Anisimov, 
Phys.\ Rev.\ B {\bf 52}, R5467 (1995).

\bibitem{superlsda}  
O. Gunnarsson {\it et al.\/}, Phys.\ Rev.\ B {\bf 39}, 1708 (1989).

\bibitem{anigun}  
V. I. Anisimov and O. Gunnarsson, 
Phys.\ Rev.\ B {\bf 43}, 7570 (1991).

\bibitem{JUDD}  
B.R.Judd, {\it Operator techniques in atomic spectroscopy},
McGraw-Hill, New York, 1963.

\bibitem{ANISOL}  
V.I. Anisimov {\it et al.\/}, Phys.\ Rev.\ B {\bf 48}, 16929 (1993). 

\bibitem{lmto}  
O.K.Andersen, Phys.\ Rev.\ B {\bf 12}, 3060 (1975).

\bibitem{lichtexchange}  
A.I. Liechtenstein {\it et al.\/}, J. Magn.\ Magn.\ Mater.\ {\bf 67}, 65 (1987). 
\bibitem{noncol}  
B. Bernu, C. Lhuillier, and L. Pierre,
Phys.\ Rev.\ Lett.\ {\bf 69}, 2590 (1992);
L. Capriotti, A. Trumper, and S. Sorella,
preprint cond-mat/9901068.

\bibitem{gebauer} 
R. Gebauer {\em et al.\/} (unpublished).

\bibitem{core} 
L.I. Johansson, F. Owman, and P. Martensson,
Phys.\ Rev.\ B {\bf 53}, 13793 (1996) 

\end{references}
\end{document}